\documentstyle[12pt,aaspp4]{article}

\received{}
\accepted{}
\journalid{}{}
\articleid{}{}
\lefthead{Lentz, Branch, \& Baron}
\righthead{Galactic $^{26}$Al Gamma-Ray Simulation}

\begin{document}
\title{Monte Carlo Simulation of the Galactic $^{26}$Al Gamma-Ray Map} 

\author{Eric J. Lentz, David Branch, and E. Baron}

\affil{Department of Physics and Astronomy, 
University of Oklahoma, 440 W. Brooks, Norman, OK 73019--0225\\
lentz,branch,baron@mail.nhn.ou.edu}

\begin{abstract}

The observed map of 1.809 MeV gamma-rays from radioactive $^{26}$Al
(Oberlack et al, 1996) shows clear evidence of a
Galactic plane origin with an uneven distribution.  We have simulated
the map using a Monte Carlo technique together with simple assumptions
about the spatial distributions and yields of $^{26}$Al sources
(clustered core-collapse supernovae and Wolf Rayet stars; low- and
high-mass AGB stars; and novae).  Although observed
structures (e.g., tangents to spiral arms, bars, and known star-forming
regions) are not included in the model, our simulated gamma-ray distribution
bears resemblance to the observed distribution.  The major difference is
that the model distribution has a strong smooth background along the
Galactic plane from distant sources in the disk of the
Galaxy.  We suggest that the smooth background is to be expected, and
probably has been suppressed by background subtraction in
the observed map.  We have also found an upper limit of $1 M_{\sun}$ to the contribution
of flux from low-yield, smoothly distributed sources (low-mass AGB stars and novae). 

\end{abstract}

\keywords{gamma rays: theory---nuclear reactions, nucleosynthesis, abundances --Galaxy: abundances}
 
\section{Introduction}

The gamma-ray created by the decay of $^{26}$Al to $^{26}$Mg was the first
discovered Galactic gamma-ray line (Mahoney et al 1982).  The $^{26}$Al
nucleus decays by positron emission to the first excited state of $^{26}$Mg,
which subsequently decays to the ground state emitting a 1.809~MeV gamma-ray.  The mean 
lifetime of $^{26}$Al, $\tau = 1.05$ x $10^6$ years, makes the 1.809 MeV gamma ray 
line an excellent tracer for newly synthesized material
released into the ISM over the last several million years.

The main production mechanism of $^{26}$Al is proton capture on
$^{25}$Mg. Astrophysical environments that can produce $^{26}$Al include hydrostatic H-burning
in the convective cores of massive stars and the H-burning shells of intermediate
mass stars, and explosive H burning in novae.  The carbon and neon rich shells
of massive stars are also a site for $^{26}$Al production both statically and
explosively.  In addition to its production, the fresh $^{26}$Al must be transported
into the ISM before it decays in order to be observable.  The explosive mechanisms present no problems, but
the transport timescale in AGB stars is of similar order to the decay timescale
causing a reduction in the amount of $^{26}$Al released into the ISM.

The first map of Galactic 1.809 MeV gamma-ray emission from $^{26}$Al was published by
Oberlack et al (1996) from COMPTEL data.  This map has a 1$\sigma$ angular
resolution of $1.6^{\circ}$, or $3.8^{\circ}$ FWHM.  The map was produced
using a Maximum-Entropy method after background subtraction.
The 1.809 MeV gamma-ray map has several important characteristics, including the
concentration of emission in the Galactic plane, a strong, irregular emission region 
toward the inner Galaxy, and a generally uneven, or clumpy, emission distribution. 
Along the Galactic plane there are several disconnected emission regions, 
some of which have been associated with O-B associations, spiral
arm tangents, and the Vela SNR.  Chen et al (1996) also identify several of
the regions with spiral arm tangents.  A recent and thorough review of 
the entire topic including observation, sources, and distribution of $^{26}$Al
can be found in Prantzos and Diehl (1996).

These observations can best be explained with sources that are spatially
concentrated and rare.  If the major sources of emission had small yields and
a smooth Galactic distribution, the emission would be quite uniform.
This is not seen in the published results (Oberlack et al 1996), which show
large gaps along the Galactic plane between emission regions.
We have therefore built a Monte Carlo
 model for the Galactic $^{26}$Al emission containing all 
potential astronomical sources.  We have also allowed the most massive stars
to form clusters that do not dissociate in the lifetime of those stars.
We have made only the simplest of assumptions about Galactic structure, an
exponential disk and a bulge.  We have not attempted to represent any specific
observed structures in the Galaxy.  All non-uniformities arise from the random
nature of the simulation.
This produces a map that, to the eye, has a strong resemblance to the
observations, with the exception of a persistent uniform background not found
in the reduced observational data.

We plan to use more statistically rigorous methods to measure
the strength of this similarity in future work, as well as to consider some
of the effects of non-uniform structure and the enhanced resolution of the INTEGRAL
observatory.

\section{Model}

Our model of Galactic $^{26}$Al 1.809 MeV gamma-ray flux uses a
Monte Carlo model of the Galaxy to generate the raw flux data (\S~\ref{monte}),
and a gaussian smoothing technique to plot the data on an equal-area
projection of the sky(\S~\ref{dens}).

\subsection{Monte Carlo Model}\label{monte}

We have  modified the Monte Carlo model of the Galaxy developed by Hatano,
Fisher \& Branch (1997a)
to study supernova visibility.  The data are generated as a series
of point source events.  For each point in the disk the radial and vertical
 positions are drawn from exponential distributions. 
 Each point in the bulge is drawn from the distribution $(R^3 + a^3)^{-1}$,
 where R is the distance to the Galactic center and $a=0.7$ kpc. 
 The age of the event is selected uniformly from a fixed
simulation length.  The $^{26}$Al yield of each event is reduced to account
for the radioactive decay and the current decay rate is computed to give the
current gamma-ray luminosity.  The luminosity is geometrically diluted
to compute the flux at the Earth.  The data are saved individually or in
bins smaller than the detector resolution ($\frac{1}{8}^{\circ}$ vs. $1.6^{\circ}$).

The computation of the flux from clusters of massive stars requires additional
steps.  Each star cluster is assigned a random size and age.
The age of the clusters is drawn from a span which is longer than the 
simulation length by the evolutionary timescale of the slowest evolving
constituent star.  With the assumption of coeval star formation, 
this allows the stellar death rates to be in equilibrium across the 
simulation length.  The mass of each star in a cluster is drawn from a power-law
initial mass function (IMF) of the form $f(m) \propto m^{-2.7}$.
 The mass of the star will determine the evolutionary
timescale and yield of the subsequent supernova.  The age of each contributing
event is computed by subtracting the evolutionary timescale from the age of
the cluster.  Stars that have not yet reached the end of their evolutionary
tracks are removed from the simulation.  The $^{26}$Al gamma-ray fluxes are
computed from the yields as described above.  If the star is a Wolf-Rayet
progenitor, the appropriate $^{26}$Al yield will be added at a fixed time 
before the end of stellar evolution.  The values of the yields, scales, etc. 
will be given in the section describing the simulation(\S~\ref{sim}).

\subsection{Flux Mapping Method}\label{dens}

We have also developed a procedure to translate the randomly placed data
points into an intensity map.  The data generated in \S~\ref{monte} 
were first sorted into bins one degree on each side to speed
later calculations.  To compute the local intensity at any point, the flux from
each point within a circular window of fixed radius was summed using a Gaussian 
weight dependent on the distance between the data point and evaluation point.
The radius of the fixed circle was chosen to be three times the smoothing
length or Gaussian width,~$\sigma$.  This value was found to be good to about
$0.1\%$ by simple tests with centrally peaked and flat test functions.  All 
distances were computed in degrees of arc, and the data bins that were used
to compute each flux point were carefully selected to cover the entire area
of the circular window.  The local flux was computed at one degree intervals
in Galactic coordinates, and plotted with contours of 
$1.57~\times 10^{-5}~\gamma$~cm$^{-2}$ s$^{-1}$ in
Figure~\ref{fig:all}.\footnote{Plots made using routines from the PGPLOT Graphics Subroutine Library by T.~J. Pearson.\label{pgplot}}

\section{The Simulation}\label{sim}

In constructing our model of the Galaxy we have chosen a bulge with radius 3~kpc,
and a disk that extends to a radius of 20~kpc with a radial scale length of 5~kpc.
The Earth is placed in the Galactic plane at a radius of 8~kpc.  Objects with
low-mass progenitors, (novae, AGB stars), have a disk component that extends to
the center of the Galaxy with a vertical scale height of 350~pc.  The low-mass
AGB stars and novae also have a bulge component with a 7:1 disk-to-bulge ratio
as per the nova implementation of the same model (Hatano et al 1997b).  
High-mass objects (SN, W-R) consist only of a disk component with a 50~pc 
scale height that does not include the inner 3~kpc of the Galaxy where the 
bulge is located.  Additionally,the high-mass objects are clustered into groups
of $10 \pm 2\sigma$ stars, where $\sigma$ is four.

\subsection{Source Frequencies}\label{sim:freq}

To compute the rate of the component sources in the model we have followed the
analysis of Prantzos and Diehl (1996) on the rate of $^{26}$Al production.
We therefore choose an initial mass function of $f(m) \propto m^{-2.7}$ 
for progenitors with $M > 1 M_{\sun}$,
and a star formation rate of $\sim 5$ stars yr$^{-1}$.  We also refer to AGB
stars with 1--4 $M_{\sun}$ progenitors as low-mass and those with 4--9 $M_{\sun}$
progenitors as high-mass.  Supernova and Wolf-Rayet progenitors will be those
stars from 9 $M_{\sun}$ up to 120 $M_{\sun}$.  For novae we have adopted the
value 40 yr$^{-1}$ suggested by Hatano et al (1997b) using the same model geometry.
Table~\ref{tab:source} summarizes the rates, yields, and model fluxes for the  
$^{26}$Al sources used to make Figure~\ref{fig:all}.

\subsection{$^{26}$Al Yields}\label{sim:y}

The yields of Weaver and Woosley (1993) are used for supernovae.  
These models use a large grid of nuclei, to give more accurate results
in the synthesis of various isotopes, in the pre-supernova phase and
the explosive phase.  The yields of Meynet et al (1997) are used for Wolf-Rayet phase sources.
These models also use an expanded network of nuclear reactions to cover the MgAl
chain.  Models were calculated for three metallicities, $Z=0.008, 0.020, 0.040$.
Our model Galaxy includes three radial zones, with inner radii of 12~kpc, 6~kpc, and
3~kpc respectively for the three metallicities.
AGB stars are divided into low- and high-mass with the division at 4 $M_{\sun}$
as in Prantzos and Deihl (1996). We follow their use of
$3 \times 10^{-5} M_{\sun}$ $^{26}$Al per high-mass AGB star from Bazan et al (1993) and 
$10^{-8} M_{\sun}$ $ ^{26}$Al per low-mass AGB star from Forestini, Paulus, \& Arnould
(1991).
For novae we use $5 \times 10^{-10} M_{\sun}$ $^{26}$Al for CO novae and
$8 \times 10^{-9} M_{\sun}$ $^{26}$Al
for ONe novae from the models of Jos\'{e}, Hernanz, and Coc (1997).
We discuss a limit to the smooth source contribution in \S~\ref{smooth}.

\subsection{Best Model}

\begin{figure}

\plotone{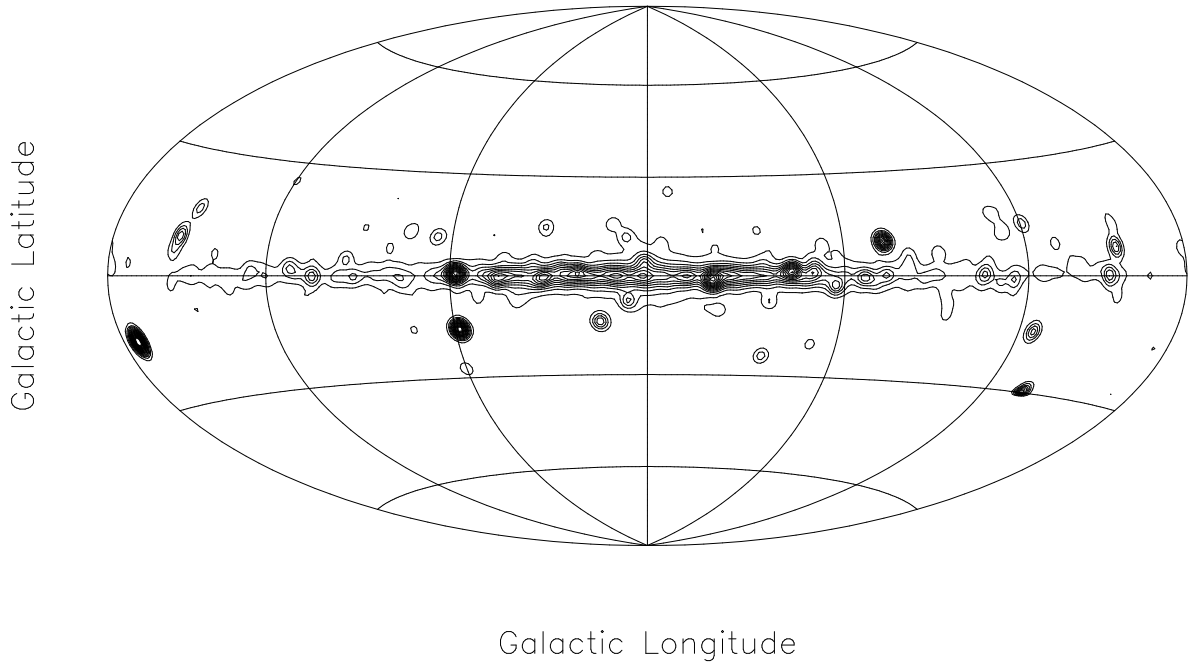}

\figcaption{Contour plot of simulated intensity using an all-sky projection
with a contour interval of $1.57~\times 10^{-5}~\gamma$~cm$^{-2}$ s$^{-1}$ 
This plot is to be compared to the observed map of Oberlack et al. (1996).
\label{fig:all}} \end{figure}

\begin{figure}
\epsscale{0.75}
\singlespace
\plotone{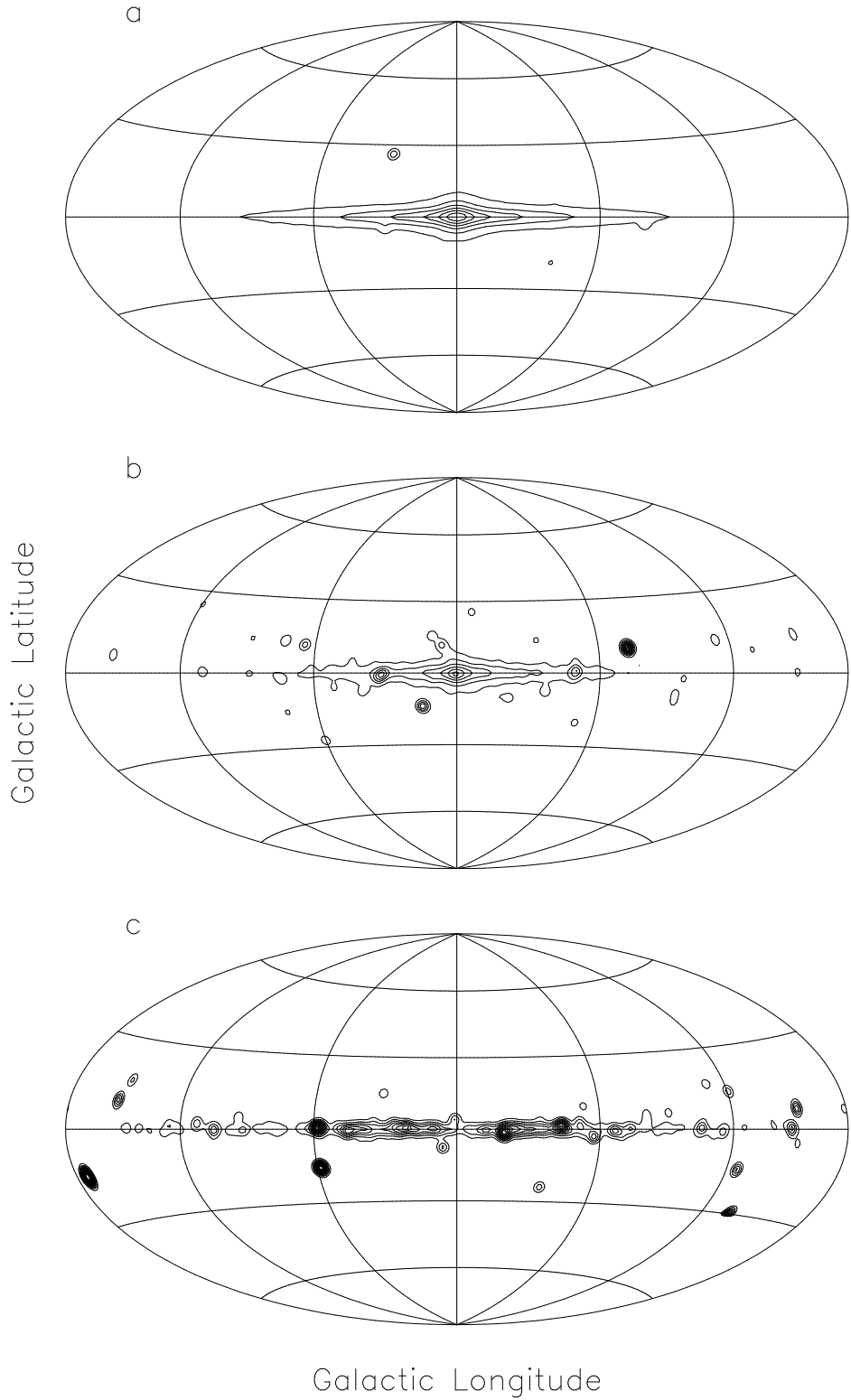}
\figcaption{Intensities of model components in Figure~\ref{fig:all} plotted with contours
of $1.57~\times 10^{-5}~\gamma$~cm$^{-2}$ s$^{-1}$. These include, a. the smooth component 
(novae and low-mass AGB stars) multiplied by 10, b. high-mass AGB stars, and
c. objects with massive star progenitors (SNe and W-R stars).\label{fig:comp}} \end{figure}

Combining the sources in with the self-consistent frequencies in \S~\ref{sim:freq} 
and the yields in \S~\ref{sim:y} gives the map in Figure~\ref{fig:all} using the same
contours as in Oberlack et al (1996) for comparison.  Figures~\ref{fig:comp}a,b,c are 
the smooth (novae and low-mass AGB stars plotted with contours one-tenth the standard value),
high-mass AGB star, and massive
star (W-R and SNe) components respectively,
used to make Figure~\ref{fig:all} using the same contours.  We see that high-mass AGB
stars provide some of the irregularity seen in the observations and that massive star
sources provide the concentration of flux from the inner region of the Galaxy,
$|\ell| < 30^o$.  The SNe/W-R component provides most of the observed irregularity.
The smooth, low-yield component (novae and low-mass AGB stars) does not make a detectable
contribution in this simulation.  The extent to which larger contributions can be made
by these sources without distorting the results in Figure~\ref{fig:all} is given in
\S~\ref{smooth}.  The total flux from each sub-component can be found in 
Table~\ref{tab:source}.  The component contributions are
0.12~$M_{\sun}$~$^{26}$Al from the smooth component, 0.93~$M_{\sun}$~$^{26}$Al from
high-mass AGB stars, and 0.76~$M_{\sun}$~$^{26}$Al from massive stars.

\subsection{Detectability}\label{sig}

For a point source the $3\sigma$ detection flux for narrow lines at 1.8 MeV with
COMPTEL, $F_{3\sigma}$, is $3 \times 10^{-5}$~$\gamma$~cm$^{-2}$ (Sch\"{o}nfelder et al 1993)
for a $10^6$ second exposure.  When incorporated with the Gaussian smoothing kernel 
(instrument response function) of $\sigma = 1.6^{\circ}$ the $3\sigma$
intensity limit, $I_{3\sigma}$, is $6 \times 10^{-3}$~$\gamma$~cm$^{-2}$~s$^{ 1}$~sr$^{-1}$
for a $10^6$ second exposure.  The data used in Oberlack et al (1996) represents 3.5 years
of COMPTEL observations with exposures along the Galactic plane from $\sim 35$ -55 $\times
10^6$ seconds.  We have chosen $45~\times$~$10^6$ seconds as the representative exposure time for the simulated
map.  Using this exposure, the  $1\sigma$ detection limit, $I_{1\sigma}$, is 
$3 \times 10^{-4}$~$\gamma$~cm$^{-2}$~s$^{-1}$~sr$^{-1}$.  This is about twice the contour
interval used in the map of Oberlack et al (1996) and in Figures~\ref{fig:all}\&\ref{fig:comp}.
These contours then approximately represent one-half sigma confidence contours.

\section {Comparison with Observations}

The main goal of this paper is to reproduce the ``look and feel" of the observed 
intensity map of Oberlack et al (1996).  It should be noted that we have made
no attempt to reproduce individual features of the observations by placing individual sources and clusters in specific places.  Such an approach would lead to an artificial
reproduction of the observation by introducing too many free parameters.  

Like the observations, the central region of the simulation, Figure~\ref{fig:all},
the inner $\pm 30^o$ shows a strong region of emission closely
restrained to the Galactic plane with internal irregularities.  As in the observations, 
the central emission region ends rapidly $30^o$ from the Galactic Center.
Disconnected peaks and emission regions along the Galactic plane outside of the inner
Galaxy appear in both obervation and simulation.

The main difference between the simulation and observation is the 
presence of an extended background along the Galactic plane at the level of the first
plotted contour in the simulation.  As noted in \S~\ref{sig}, the first plotted contour is roughly 
equivalent to one-half sigma confidence.  We suspect that the complex and difficult
extraction of the 1.8 MeV gamma-line map from the significant background has resulted in the disappearance
of the low-level background. We also suspected the maximum entropy method (MEM) 
may be responsible for suppressing the low-level background, but the
application of MEM to the simulated data did not cause the low-level background to disappear.
This leads us to suspect that the background subtraction is the likely
cause for the disappearance of the low-level background from the observations. 

Only points within the second contour in our simulations
would have even a 2/3 chance of being observed in the current COMPTEL data.  We 
suggest that with the sensitivity to see $10^{-4}$~$\gamma$~cm$^{-2}$~s$^{ 1}$~sr$^{-1}$ intensities with $1\sigma$
confidence or better, about $400 \times 10^6$ seconds of exposure, a low-level background
along the plane from distant and indistinguishable sources would be inevitably
found.

\subsection{Limit on Smoothly Distributed Sources} \label{smooth}

\begin{figure}
\plotone{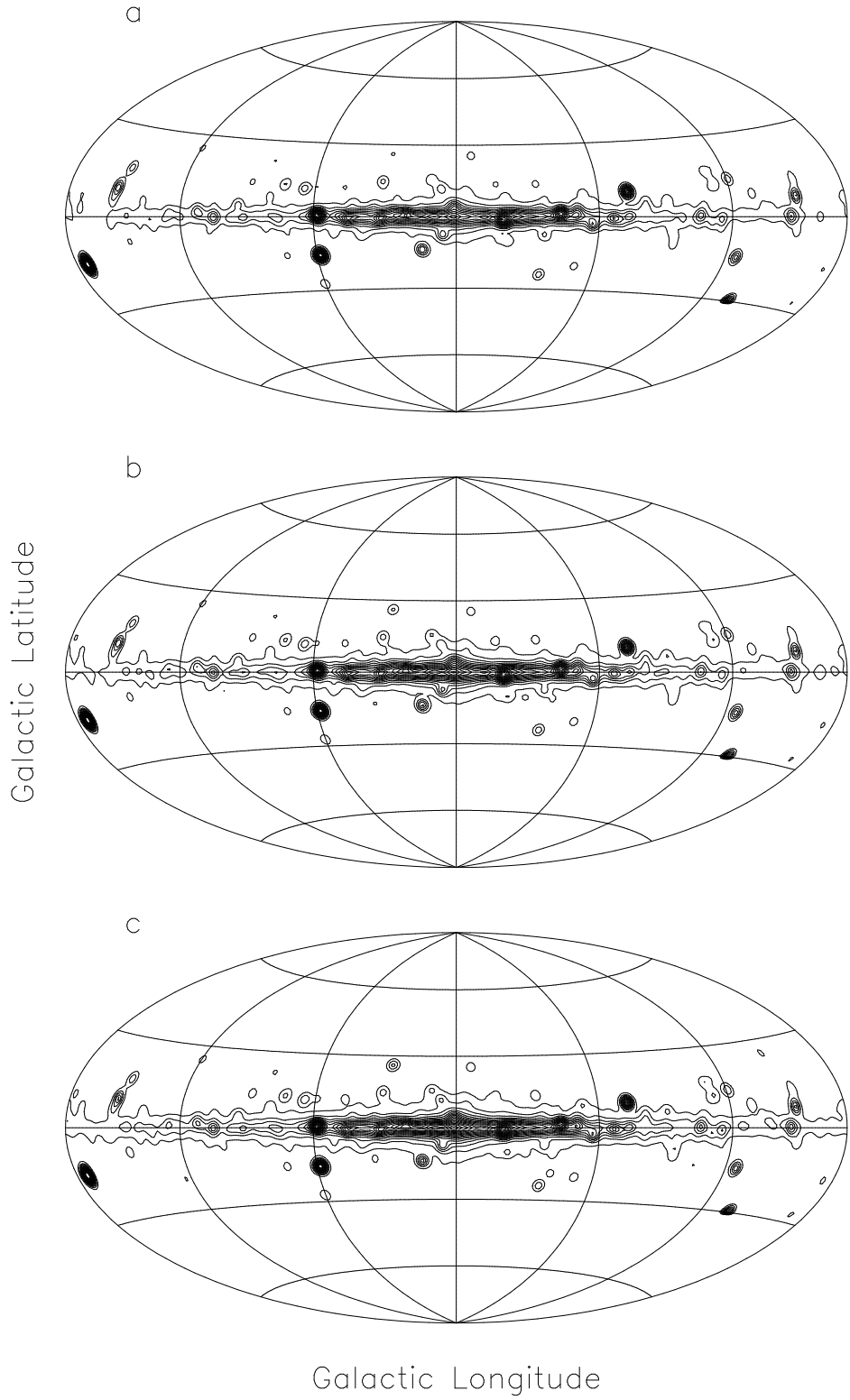}
\figcaption{Contour plot of simulated intensity using $1.57~\times 10^{-5}~\gamma$~cm$^{-2}$ s$^{-1}$ 
contours with a. 4 times, b. 8 times, c. 12 times
the smooth source component in Figure~\ref{fig:all}.\label{fig:smooth}}\end{figure}

The ``smooth" component, low-mass AGB's and novae (Figure~\ref{fig:comp}a), contains
about 1/8~$M_{\sun}$ of $^{26}$Al which emits about 
$5 \times 10^{-5}~\gamma$~cm$^{-2}$~s$^{-1}$, or about 1\% of the total flux.  We 
have tested how much of this smooth component can be added to the remaining
sources without distorting the sum from the best model (Figure~\ref{fig:all}) to
something incompatible with the map in Oberlack et al (1996).  We have produced
three models shown in Figures~\ref{fig:smooth}a,b,c that contain 4, 8, and 12 times
the smooth component of our regular model respectively.  The 4-fold model 
(Figure~\ref{fig:smooth}a, 0.5 $M_{\sun}$~$^{26}$Al) has only
minute differences from Figure~\ref{fig:all},
that require the two plots to be overlaid to be seen.  The 8-fold model 
(Figure~\ref{fig:smooth}b, 1 $M_{\sun}$~$^{26}$Al) shows a thickening of the inner
Galaxy emission region and a small extension of the $1\sigma$ (second contour) 
background outside the inner Galaxy which is acceptable, but near the limit where it
would be detectable with the current observations.  The last
model, Figure~\ref{fig:smooth}c with 1.5 $M_{\sun}$~$^{26}$Al, shows considerable
thickening of the inner Galaxy emission and an unacceptably long extension of
Galactic plane background at the $1\sigma$ detection level.  If such a large background
were present, the Oberlack et al (1996) analysis should have shown stronger evidence
of its existence.  Therefore despite uncertainties in yields and rates
for all objects, about 0.5 $M_{\sun}$~$^{26}$Al can be easily hidden in the smooth
background without detection or modifying the results of this simulation, with an
upper limit of about 1.0 $M_{\sun}$~$^{26}$Al in the smooth background.

\section {Conclusions}

Our simulation has shown that as expected by previous authors (e.g., Prantzos \& Diehl, 1996),
the sources with massive progenitors make most of the flux and provide
the irregular structure seen in the observed map (Oberlack et al 1996).  We have 
found that a background of strong sources diluted by distance should be detectable, 
with a longer exposure time.  While the source rates and yields are uncertain we
can limit the amount of $^{26}$Al generated by the frequent but small `smooth' sources
to be about 1.0 $M_{\sun}$.  The total mass of $^{26}$Al for the high-yield, massive 
progenitor sources is about 1.7 $M_{\sun}$.  This model produces a reasonable 
approximation of observation.  The rates and yields of the massive progenitor sources 
are also uncertain, but changes in yields can be compensated for by changes in rates
as long as the number of distinct emission sites (clusters or high-mass AGB stars)
does not change by a large factor.

This paper uses a Monte Carlo code of the Galaxy derived from the original
version written by Adam Fisher.
This work has been supported in part by NSF grants AST 9417102 and 9417242 and
NASA grant NAG5-3505.
\clearpage
\begin{deluxetable}{lccclr}
\footnotesize
\tablecaption{Sources of $^{26}$Al in Model. \label{tab:source}}
\tablewidth{0pt}
\tablehead{\colhead{Source} &\colhead{Rate}   &\colhead{Yield} &\colhead{Model Flux} \\
& \colhead{(cen$^{-1}$)} &\colhead{($M_{\sun}$)} &\colhead{($\gamma$ cm$^{-2}$ s$^{-1}$)} }
\startdata
Novae (CO)  &2800 &$5 \times 10^{-10}$ &$3 \times 10^{-7}$ &bulge \nl
  & & &$ 6 \times 10^{-6}$ &disk \nl
Novae (ONe) &1200 &$8 \times 10^{-9}$ &$2 \times 10^{-6}$ &bulge \nl
  & & &$ 4 \times 10^{-5}$ &disk\nl
Low-Mass AGB &40 &$10^{-8}$ &$10^{-7}$ &bulge \nl
  & & &$ 2 \times 10^{-6}$ &disk \nl
High-Mass AGB &3 &$3 \times 10^{-5}$ &$5 \times 10^{-4}$ \nl
Massive Stars & & &$3.2 \times 10^{-3}$ \nl
Supernovae &1 &$10^{-5}$ to $10^{-4}$ & \nl
Wolf-Rayet\tablenotemark{\dagger} & &$10^{-5}$ to $10^{-3}$ \nl
\enddata
\tablenotetext{\dagger}{Wolf-Rayet rate appropriate to related supernova progenitor}
\end{deluxetable}

\clearpage

\clearpage

\clearpage

\end{document}